\documentclass[epjST]{svjour}
\usepackage{graphics}\usepackage{graphicx}\usepackage{wrapfig}
\usepackage{sidecap}
\usepackage{amstext} \usepackage{amsmath} \usepackage{amssymb}

\begin{document}
\title{Designing spin-spin interactions with one and two dimensional ion crystals in planar micro traps}
\author{J. Welzel \inst{1}\fnmsep\thanks{\email{welzel@uni-mainz.de}} \and A. Bautista-Salvador\inst{1} \and C. Abarbanel \inst{2} \and V. Wineman-Fisher \inst{2} \and C. Wunderlich \inst{3} \and R. Folman \inst{4} \and  F. Schmidt-Kaler\inst{1}}
\institute{\inst{1} Johannes Gutenberg Universit\"at Mainz, Institut f\"ur Physik, QUANTUM, 55128 Mainz, Germany\\ \inst{2} Ilse Katz Institute for Nanoscale Science and Technology, Ben-Gurion University of the Negev, Be'er Sheva 84105, Israel\\ \inst{3}  Universit\"at Siegen, Quantenoptik,  57068 Siegen, Germany\\ \inst{4} Department of Physics, Ben-Gurion University of the Negev, Be'er Sheva 84105, Israel}
\abstract{We discuss the experimental feasibility of quantum simulation with trapped ion crystals, using magnetic field gradients. We describe a micro structured planar ion trap, which contains a central wire loop generating a strong magnetic gradient of about 20~T/m in an ion crystal held about 160~$\mu$m above the surface. On the theoretical side, we extend a proposal about spin-spin interactions via magnetic gradient induced coupling (MAGIC) [Johanning, et al, J. Phys. B: At. Mol. Opt. Phys. \textbf{42}, (2009) 154009]. We describe aspects where planar ion traps promise novel physics: Spin-spin coupling strengths of transversal eigenmodes exhibit significant advantages over the coupling schemes in longitudinal direction that have been previously investigated. With a chip device and a magnetic field coil with small inductance, a resonant enhancement of magnetic spin forces through the application of alternating magnetic field gradients  is proposed. Such resonantly enhanced spin-spin coupling may be used, for instance, to create Schr\"odinger cat states. Finally we investigate magnetic gradient interactions in two-dimensional ion crystals, and discuss frustration effects in such two-dimensional arrangements.} 
\maketitle
\section*{Introduction}
\label{intro}
When many atoms of equal or different species bind together and form a piece of matter, its properties depend on the mutual interactions of the constituents and the structure which was formed. Surprising collective phenomena appear in the magnetic properties of a lattice, for example when spins are aligned parallel or anti-parallel with each other, as observed in ferromagnetic or anti-ferromagnetic materials, respectively. Also the electric transport properties in such materials can undergo phase transitions, or insulators or superconductors may form, depending on control parameters such as chemical composition, dopant concentrations, pressure, temperature and magnetic field. The goal of predicting such material properties from first principles and even to design these properties in view of applications is difficult because the complexity of quantum interactions increases rapidly with the number of particles. Here, the importance of quantum optical model systems becomes apparent, as one can construct a complex system from individual, fully controlled atoms or ions, where even the mutual interactions are tunable by external control parameters over a large range. This bottom-up approach of simple model systems promises to allow a verification of the computational methods, as well as the investigation of the basic properties of quantum magnetism and superconductivity. A long term goal is thus the simulation of real material systems with such quantum optical model systems, following the vision of R. Feynman who, in 1982, proposed simulating the static and dynamical properties of systems that are difficult or even impossible to solve numerically by investigating experimentally a well controlled model system \cite{Feynman82}. This upcoming field has been attractive for both experimental and theoretical physicists \cite{Barreiro10,Weimer10}.

Experiments with trapped atoms in optical lattices have pioneered the field, following the proposal by Jaksch, Cirac and Zoller \cite{Jaksch05,Cirac00,Cirac95} with spectacular demonstrations of the Mott insulator to superconducting phase transition \cite{Bloch2008}. Today these experiments aim for complex lattice structures, including superlattices, frustration in Kagome arrangements \cite{Sen2008,Santos2004}, or a simulation of SU(N) magnetism \cite{Gorshkov10,Taie10}. Trapped ion quantum simulation has been proposed by Porras and Cirac \cite{Porras04}, where optical dipole forces from laser beams induce spin dependent forces to generate effective spin-spin interactions. Experimentally, we have seen the simulation of a magnetic quantum phase transition with two ions \cite{Schaetz08} and more recently the simulation of a frustrated three-spin anti-ferromagnetic interaction with three ions \cite{Kim10}. Designed Hamiltonians may be used to simulate the Dirac equation and relativistic Hamiltonians \cite{ROOS2010a,GER2010}, or crystal-to-superfluid transitions in XXZ-spin chains \cite{Hauke10}. Yet another way to design spin-spin interactions has been proposed by Wunderlich \cite{Wunderlich02}, where strong static magnetic gradients are used. In an  inhomogeneous magnetic field, spins experience forces such that the ions are displaced from their equilibrium positions in the ion crystal. This leads to a mutual effective spin-spin interaction. First experiments with a single ion have demonstrated the spin-dependent force \cite{Wunderlich09}; the generation of many-particle entangled quantum states \cite{WunderlichPRA} and a generalization with rf-magnetic fields \cite{Osp08,Johanning09} has also been proposed. As the interactions are generated without laser light fields, but only by magnetic gradients, scaling of the simulation to more particles might be facilitated \cite{Chiaverini08}.

We describe in this paper how magnetic interactions may be simulated when a cold crystal of ions in a planar ion trap is used. We discuss the specific advantages of our approach in a particular two-dimensional trap, where ions come close to the surface, whereby they may experience large magnetic gradients which are necessary for significant spin-spin interaction strengths. Furthermore, the planar trapping device should permit the arrangement of ions for a quantum simulator either with ions in one linear crystal or in multiple parallel linear crystals, thus forming a two-dimensional structure. In Sec.~\ref{planar_trap} we will therefore summarize the basics of planar traps and describe the specific features of the device we have developed. We estimate the strength of the expected magnetic field gradients in Sec.~\ref{mag_field}.  Sec.~\ref{sterngerlach} then shows how spin-spin interactions are generated and how the strength can be designed depending on the applied magnetic field gradients and their direction, allowing one  to switch from a ferro- to antiferromagnetic interaction. Sect.~\ref{AC_Current} describes how the time dependence of the corresponding current  resonantly selects specific eigenmodes of the crystal. A scheme for a two dimensional ion crystal well suited for quantum simulations is presented in Sec.~\ref{2D}, as well as the complete calculation of eigenmodes and spin-spin interaction strengths in Sec.~\ref{ferroantiferro}. All parameters given in the paper are within reach of our on-going experiment using the particular planar trap \cite{BGU} also described in Sect.~\ref{planar_trap}. The goal of this paper is to highlight possible avenues towards the quantum simulation of magnetic systems with ion crystals and to stimulate the discussion of further types of spin interacting multi-particle systems.

\section{Planar traps with magnetic field gradients}
\subsection{Planar micro traps} \label{planar_trap}
\begin{figure}[]
\begin{minipage}[hb]{0.5\textwidth}
\centering
\includegraphics[width = 0.865\textwidth]{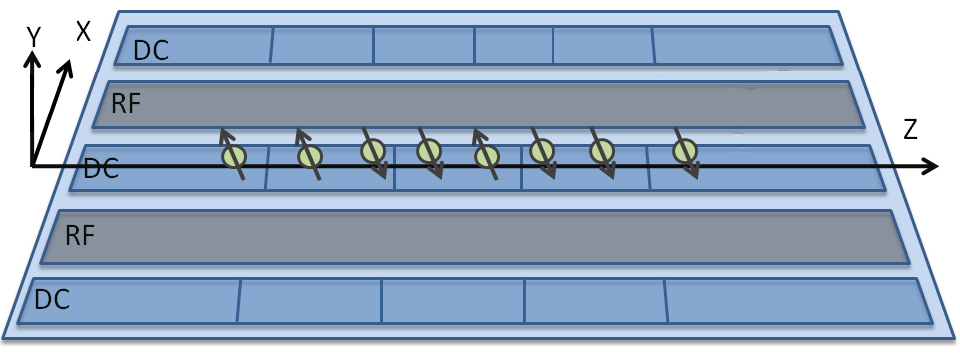}
\end{minipage}
\begin{minipage}[hb]{0.5\textwidth}
\centering
\caption{Typical arrangement of the electrodes in a planar trap. A linear trapped ion crystal is sketched oriented along the z-axis.}
\label{SkizzePT}
\end{minipage}
\end{figure}
Using static and dynamic electric fields to hold ions, the planar trap evolved from the linear Paul trap \cite{Chiaverini05}. Contrary to the latter, where the electrodes are configured in three dimensions, in a planar trap they are all aligned in one plane and the ions are trapped above the surface housing the electrodes. This adaptation from its previous form allows easier microfabrication using advanced lithographic techniques, which is especially important when aiming for the complex structures required for scalability towards large numbers of ions \cite{Kielpinski02}. On the other hand, in planar traps, where the ions are typically $40-200~\mu$m above the conducting electrodes, the potential is shallow and the close vicinity to the metallic surface may lead to heating \cite{Daniilidis10,Seidelin06,Labaziewicz08}. Most of the up-to-date realizations utilize five parallel electrodes \cite{Wang09,Amini08} (see Fig.~\ref{SkizzePT}). Two of them are driven by radio frequency (rf) voltages creating a quadrupolar dynamical trapping field in $x$ and $y$, while the others are held at static voltages (dc) which confine ions in the $z$ direction.

\begin{figure}[]
\begin{minipage}[hb]{0.48\textwidth}
\centering
\includegraphics[angle=180,width = 0.865\textwidth]{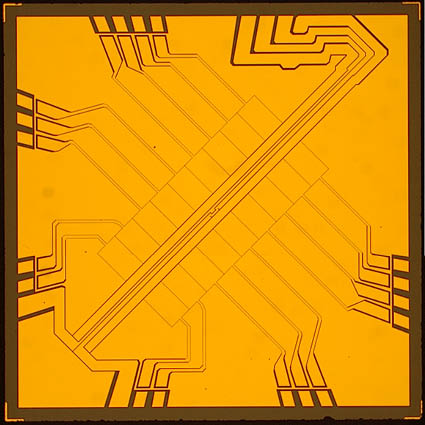}
\end{minipage}
\hfill
\begin{minipage}[hb]{0.48\textwidth}
\centering
\includegraphics[width = \textwidth]{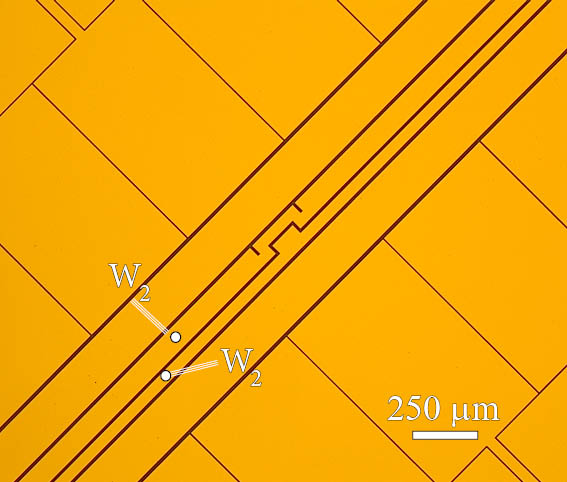}
\end{minipage}
\caption{Left: Top view of the planar trap with optical microscopy \cite{BGU}. The 5 wire configuration with dc, rf, dc, rf, dc, is modified with the center dc electrode divided into two wires $W_1$ and $W_2$ and a nine-fold segmentation of both outer dc electrode. The whole chip has a width of about 10~mm. Right: Zoom of the center of the trap. The outer dc electrodes have a width of 595~$\mu$m, the rf electrodes of 197~$\mu$m and the inner dc electrode $W_1$ a width of 100 $\mu$m. The gold layer is 1.9$~\mu$m thick. The gap between rf electrodes and their neighbors is $12~\mu\text{m}$ to avoid electrical breakdown when applying high voltages. A gap of  $3~\mu\text{m}$ is chosen between dc electrodes.}
\label{GIF}
\end{figure}

The resulting potential for a particle with charge $q$ and mass $m$ is $\phi(x,y,z,t) = \phi_{\mathrm{dc}}(x,y,z) +\phi_{\mathrm{rf}}(x,y,z) \cos{\Omega t}$ leading to an equation of motion given by
\begin{eqnarray}\label{Motionequation2}m\frac{d^2 x_i}{dt^2}+q\sum_{j}\left[\left(\frac{\partial^2\phi_{dc}}{\partial x_{i}\partial x_{j}}\right)+\left(\frac{\partial^2\phi_{rf}}{\partial x_{i}\partial x_{j}}\right)\cos(\Omega t)\right] x_{j} = 0,
\end{eqnarray} where $\Omega$ is the angular frequency of the rf potential, $x_1=x,~x_2=y,~ x_3=z$ and the potential is expanded to second order around the equilibrium position. Rewritten in standard Matthieu form, the equation of motion is
\begin{eqnarray}\label{Motionequation3}m\frac{d^2 \textbf{x}}{d\tau^2}+\left[A+2Q\cos\left(2\tau\right)\right] \textbf{x} = 0,
\end{eqnarray}
where
\begin{eqnarray}\label{AQWerte}A_{ij}=\frac{4q}{m\Omega^2}\left(\frac{\partial^2\phi_{dc}}{\partial x_{i}\partial x_{j}}\right), ~~~Q_{ij}= \frac{2q}{m\Omega^2}\left(\frac{\partial^2\phi_{rf}}{\partial x_{i}\partial x_{j}}\right),
\end{eqnarray} are the entries for matrices A (stiffness matrix) and Q (excitation matrix) determining the stability parameters of the trap \cite{House08} and $\tau =\Omega t/2$. The trajectory solution of the ion undergoes oscillations including both the so-called micromotion frequency at $\Omega/2\pi$ and the secular motion frequencies at $\kappa_{i}\Omega/2$, where $\kappa_i \approx \sqrt{A_{ii}+Q_{ii}^2/2}$ in the case that the matrices are diagonal and the entries are much smaller than one. For the accurate numerical calculation of the trap potential without approximations we employ a numerical toolbox \cite{Singer10} reaching a $1~\%$ accuracy \cite{Huber10}. With the geometry of the present chip (see Fig.~\ref{GIF}) we calculate that the trap center is $y_0  = 164~\mu \text{m}$ above the surface. Along the z-axis the segmented dc electrodes allow trapping both above the center of the winding inner electrode and also at a distance of more than $\pm$~1000~$\mu$m. Using an rf amplitude of $V_ \mathrm{{rf}} = 230~\text{V}$ at a frequency of $\Omega =2\pi  \cdot 34.88~\text{MHz}$ the depth of the trapping potential for  $^{40}\text{Ca}^+$ is $150~\text{meV}$ and the trapping frequencies are $\omega_x= 2\pi\cdot3.14~\text{MHz}$ and $\omega_y = 2\pi\cdot 3.20~\text{MHz}$ in the radial directions, and $\omega_{z}= 2\pi\cdot310~\text{kHz}$ in the axial direction (see Tab. 1 in Appendix B).

\begin{figure}[]
\begin{minipage}[hb]{0.47\textwidth}
\centering
\includegraphics[width = 0.865\textwidth]{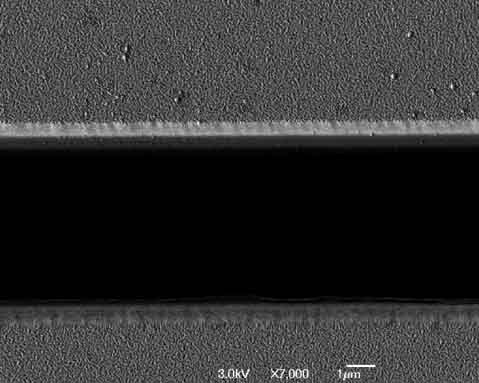}
\end{minipage}
\begin{minipage}[hb]{0.47\textwidth}
\hfill
\caption{Scanning electron microscope picture showing the quality of the insulating gap between the gold electrodes. The scale bar is 1~$\mu$m long. The surface roughness has
an RMS value of less than 5~nm, and
a peak-to-peak edge roughness  of 40~ nm. The gold resistivity is 2.63~$\mu\Omega$cm.}
\label{SEM}
\end{minipage}
\end{figure}
For the fabrication of such devices, various methods may be used. Photolithography, electroplating and etching on Borosilicate glass has led to a first generation of traps. For the details of the fabrication process we refer to Ref.~\cite{Hellwig10}. A newly established process builds on the good heat conductivity and surface quality of a sapphire substrate, on which a photolithographic procedure is applied. The developed resist serves as a mask for the deposition of a $50~\text{nm}$ thick adhesion layer of titanium and a high quality gold film of about 2~$\mu$m thickness, both being deposited by a thermal evaporator. The Ti and Au deposited on the resist top are removed in a lift-off process. From an AFM analysis we find an rms surface roughness of less than $5~\text{nm}$, while an SEM analysis determines the edge roughness to  be 40 nm (peak-to-peak) (see Fig.~\ref{SEM}). The process has an overall alignment accuracy of better than 1~$\mu$m. In our tests, we continuously apply 205~V amplitude without observing any sparking which we attribute to the high quality of the electrodes without sharp edges.
\begin{figure}
\centering
\resizebox{0.85\columnwidth}{!}{
\includegraphics{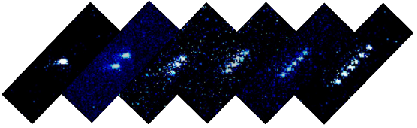} } \caption{Trapped linear ion crystals of  Ca$^+$. The applied radio frequency at 30.82~MHz with a V$_{pp}$=400~V and the dc voltages of 2.2, 0.7, -0.8, 0.7, 2.2~V  on five adjacent pairs of electrodes result in an axial frequencies of $\omega_{z}$ = 2$\pi$ 234~kHz.}
\label{ionenbild}
\end{figure}

\subsection{Trapped linear ion crystals \label{trapped_crystals}}
The above structure has been loaded with linear ion crystals of $^{40}$Ca$^+$, see Fig.~\ref{ionenbild}. We load the ions by a photoionization with laser light near 423~nm and 375~nm. The ions are cooled on the dipole transitions near 397~nm and 866~nm. The emitted fluorescence is imaged on a EMCCD camera.

\subsection{Integrated magnetic field gradient coils \label{mag_field}}
\begin{figure}[hbt]
\centering
\resizebox{0.85\columnwidth}{!}{
\includegraphics{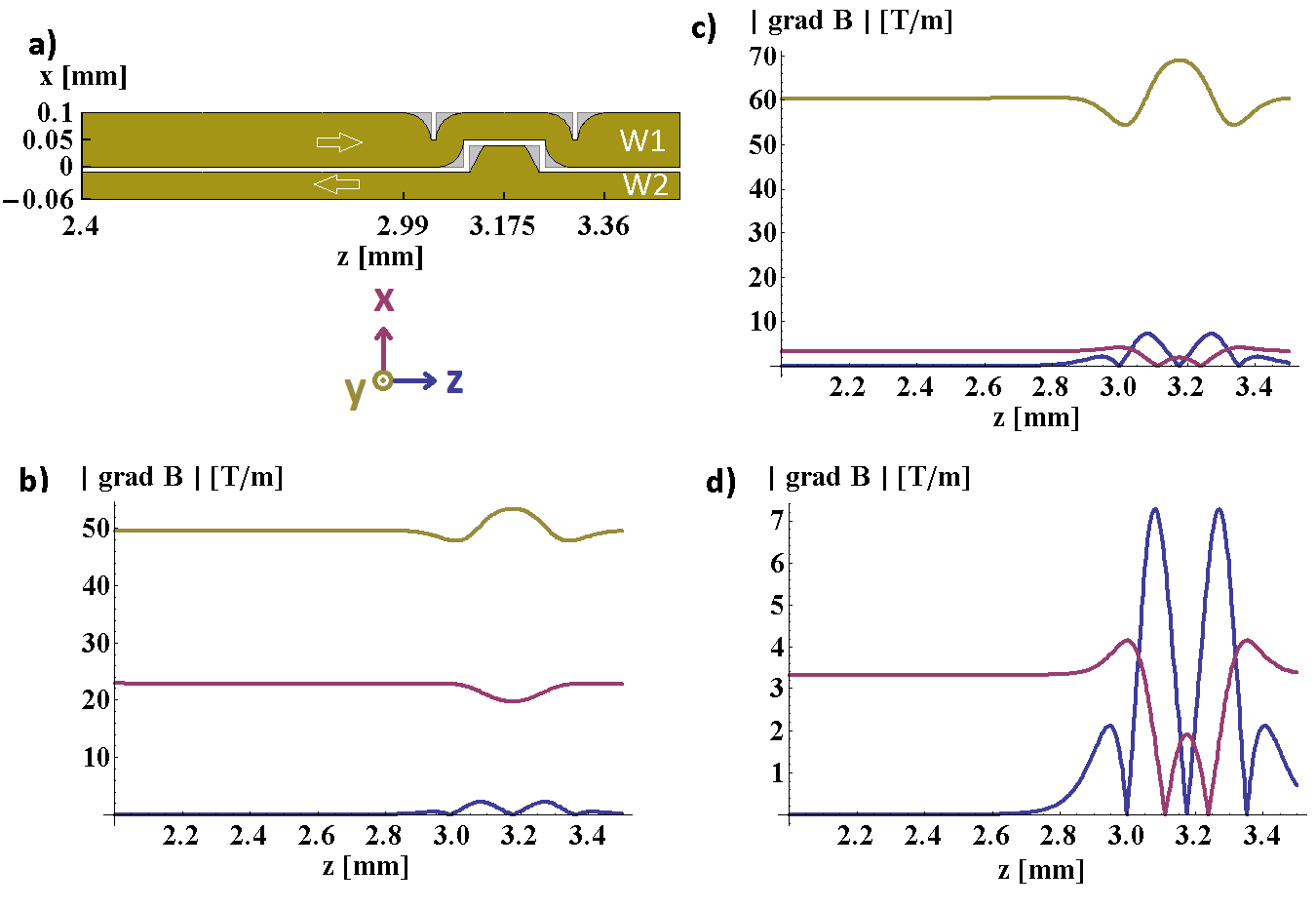} } \caption{Magnetic gradient $\bigtriangledown |\vec{B}|$ in $x$, $y$ and $z$ direction.  The subplot a) shows the  current flow within the two wires $W_1$ and $W_2$. b) The upper limit for the transversal gradient in our trap is reached with $I_{W_1}=4$~A and $I_{W_2}=-10$~A. c) $I_{W_1}=-I_{W_2}=10$~A results in a higher axial gradient, which is shown magnified in subplot d). }
 \label{gradBall}
\end{figure}
The trap design includes a split inner electrode with wires W$_{1}$ and W$_{2}$  (see Fig.~\ref{GIF}). The small loop in W$_{1}$ with inner length $150~\mu$m and outer length $250~\mu$m  generates a high gradient of the magnetic field magnitude at the location where the ion crystal is trapped above the surface. For the applications discussed below, one may require a magnetic field gradient along the trap axis, preferably high and constant over the few 100~$\mu$m extent of an ion crystal. For the simulations \cite{radia1,radia2} we assume a piecewise homogeneous current density according to the geometry of the center electrode(s). The results of our numerical simulations, presented in Fig.~\ref{gradBall}, show the magnetic gradient components plotted as function of $z$. The gradient component in $z$ direction peaks with a width of about $100~\mu$m at two locations at a distance given by the loop geometry. The transversal $x$-component of the gradient becomes constant, when moving the ion along the trap axis out of the chip center. For a  current of  $I_{W_1}=4$~A and $I_{W_2}=-10$~A  one can achieve a much higher value, up to 23 T/m at a magnetic field strength about 7.6 mT. The $y$-component is even higher with more than 50~T/m.

\begin{figure}[hb]
\begin{minipage}[hb]{0.6\textwidth}
\centering
\includegraphics[width = 0.865\textwidth]{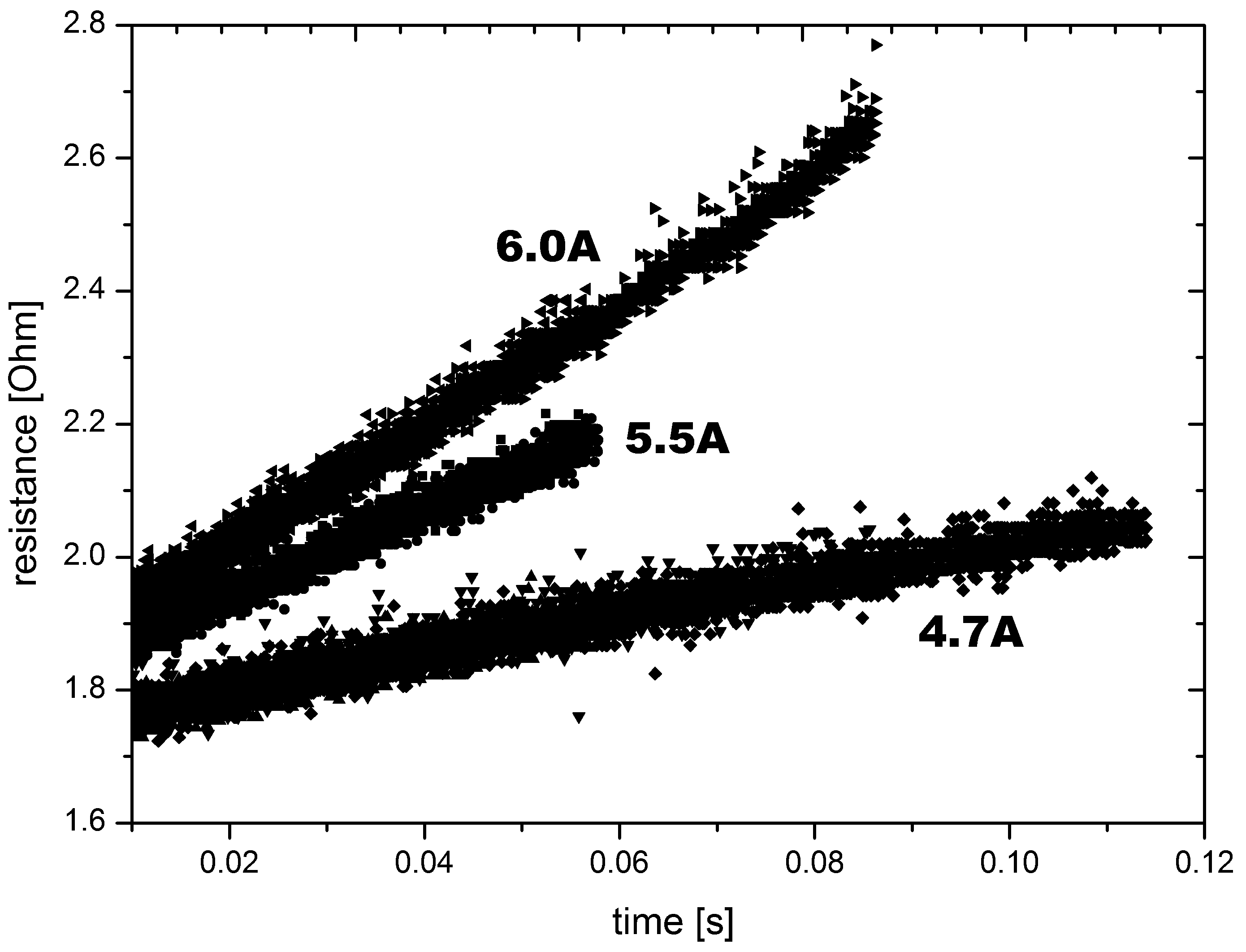}
\end{minipage}
\begin{minipage}[hb]{0.4\textwidth}
\centering
\caption{Resistance increase over time, when applying a current pulse to the center electrode $W_1$.}
\label{current}
\end{minipage}
\end{figure}

In the experiment, the magnetic field gradient will be switched on only for a short time such that the resistive heat load is reduced. In Fig.~\ref{current} we present current tests with the first generation of traps. The ratio of the measured voltage drop over the trap to the current determined via the voltage drop over a shunt is plotted as an increasing resistance over time. For a current of 6~Ampere, corresponding to a current density of $4\cdot10^6$~A/cm$^2$, we are able to apply the load for more than 0.1~s. For larger currents, we observe a faster resistance increase since the resistance of the gold wire on the chip increases with temperature. It has been found that this ultimately leads to melting of the structure \cite{Groth04}. For transmitting such high currents, we use multiple bonds where up to 10 bonding wires of diameter 25~$\mu$m are used in parallel. Our tests show that only then the gold layer on the chip is destroyed before the bonding wires break. In future, one might improve the cooling of the sapphire substrate, e.g. by a stack of Peltier elements, for even higher current densities: reducing the temperature by $\delta T \sim 100~K$ would increase the gold conductivity by more than $30~\%$. It is reported that only an increase of more than $50~\%$ in resistance will lead to  permanent damage of the conductor \cite{Groth04}. For the planned experiments one would require a few ms spin-spin interaction times, such that more than 10~A appears within reach. A high current in the central conductor goes along with a certain voltage drop, which will have an impact on the z-position of the ion crystal if not counteracted by a suited voltage bias on the current supply.

\section{Spin-spin interactions with ion crystals via the Stern-Gerlach effect \label{sterngerlach}}
Ions of mass $m$ and charge $q$ are trapped in the combined dc and ac electric potential resulting in secular motional frequencies $\omega/(2\pi)$ of typically a few MHz. With $N$ ions bound in one crystal $3N$ modes of vibrations exist, and we will review this theory in the first part of this section. If the spin-1/2 particles are exposed to a magnetic gradient, the position of the ions will change slightly due to a new force balance between the magnetic gradient-induced force on the spins and the electric restoring force of the harmonic trap potential. For a multi-ion crystal this results in an effective spin-spin interaction between the n$^{th}$ and the m$^{th}$  ion, denoted by $J_{n,m}$ \footnote{The MAGIC interaction employs Stern-Gerlach forces, but for a multi-ion crystal. Similar magnetic forces on a single electron or positron stored in a Penning trap were used for detecting flips of its spin orientation as in the seminal experiments by H. Dehmelt for the measurements of the g-factor}. The weaker the harmonic restoring force, the more the ion positions are modified by the applied magnetic gradient and the larger are the mutual spin-spin interactions. We recapitulate briefly the interactions for a magnetic gradient along the $z$-axis which result in interaction strengths typically well below 1~kHz. Aiming for stronger couplings for quantum simulations would require significantly stronger magnetic gradients. In this context we emphasize  the usefulness of transversal magnetic gradients on linear crystals \cite{Osp08,Zhu06}, since we find  that the spin-spin interactions largely increase when the trap control voltages are tuned such that the ion crystal is close to a structural  transition instability. We show that under these conditions, the corresponding spin-dependent forces even overcome the trap restoring force. The resulting interactions $J_{n,m}$ are larger by orders of magnitude as compared with axial coupling strengths. In this case the ion crystal will rearrange, contradicting the initial assumptions of the calculation. The last part of this section outlines time dependent magnetic gradients with a current $I(t)=I_{0} \sin(\omega t)$, where a resonant enhancement is employed to increase and design the couplings $J_{n,m}$.

The total potential is the sum of the trap potential, approximated by a harmonic shape of the ions excursions $x_{n,i}$ about their equilibrium  positions in cartesian coordinates $i$ and the electrostatic mutual Coulomb repulsions between the ions in the linear string: \begin{equation}
\label{eqn_pot}
V = m\omega_z^2\bigg(\sum_{n=1}^{N}\sum_{i=0}^2\frac{1}{2\alpha_i}x_{n,i}^2+\frac{l^3}{2}\sum\limits_{n=1}^{N}\sum\limits_{\substack{m=1\\ m\neq n}}^{N}\frac{1}{|\mathbf x_n - \mathbf x_m|}\bigg)\end{equation}
where $l^3 = q^2/4\pi\epsilon_0 m\omega_z^2$ denotes a scaling factor and  $\alpha_i = \omega_z^2/\omega_i^2$ denotes the anisotropy between the axial and radial confinement. The Hessian matrix $M$ (see Apendix~\ref{Hesse}) determines all motional degrees of freedom: The eigenvalues of M are the squares of the motional frequencies  $2\pi\nu_n$ of the ions, while eigenvectors of M indicate the directions of the ion motion. As long as one stays below a critical value of $\alpha_i$, ions align in a linear string and M reduces to a block diagonal matrix \cite{James98,Enzer00} with entries $m\omega_z^2 (B^x,B^y,A)$. The eigenfrequencies of $B^i$ can be expressed by those of $A$: $\sqrt{1/\alpha_i+1/2- (2\pi\nu^A_n)^2/2}$. As one can see, there is a critical value for the anisotropy, given through $\alpha_\mathrm{{crit}} = 2/((2\pi\bar\nu)^2-1)\approx 2.53N^{-1.73}$ \cite{Schiffer93}, where $\bar\nu$ is the highest eigenfrequency of Matrix $A$. When crossing this point, the  lowest eigenfrequency of $B^i$ becomes imaginary and the linear chain is no longer stable. By increasing $\alpha$ further the ions will rearrange in a zig-zag configuration to minimize the electrostatic energy.

With an additional magnetic field gradient the Hamiltonian in one dimension for a linear string of $N$ ions is given by  \cite{Wunderlich02}:
\begin{equation*}
H = \frac{\hbar}{2}\sum_{n=1}^{N}\omega_n(\mathbf{\bar x}_{n}) \sigma_{z,n}+2\pi \hbar\sum_{n=1}^{N} \nu_{n} a_{n}^\dagger a_{n}-\frac{\hbar}{2}\sum_{n<m}J_{n,m}\sigma_{z,n}\sigma_{z,m}.
\end{equation*}
The first term describes the internal energy of a two level system due to the Zeeman shift $\hbar\omega_n/2$ for the $\text{n}^{th}$ ion at its equilibrium position $\mathbf{\bar x}_{n}$ via the Pauli matrix $\sigma_{z,n}$. The confinement in the electric potential leads to vibrational eigenmodes with
frequency $\nu_{n}$. The last part represents a pairwise spin-spin coupling  between the ions: $$J_{n,m} = \frac{\hbar}{2}\frac{\partial\omega_n}{\partial x_{n,i}}\bigg|_{\bar x_{n,i}}\frac{\partial\omega_m}{\partial  x_{m,i}}\bigg|_{\bar x_{m,i}}(M  ^{-1})_{n,m}$$ determines the coupling strength using the Hessian M of the potential. The reason for such spin dependent forces is that the position of ion $m$ is slightly modified when flipping an ion's spin, leading to changes in the Coulomb repulsion energy on ion $n$, and therefore changing its position also. Ion $m$ experiences now a different Zeeman energy. Positive values of $J$  result in an energy reduction for parallel spin orientations, whereas negative values lead to reduced energy for anti-parallel spins. Decoherence effects are expected from fluctuations of the magnetic fields and might be reduced by spin echo sequences as low frequency Fourier components typically dominate the noise-spectrum.
\vspace{-0.5cm}
\begin{figure}[hbt]
\begin{minipage}[hb]{0.6\textwidth}
\centering
\includegraphics[width = 0.865\textwidth]{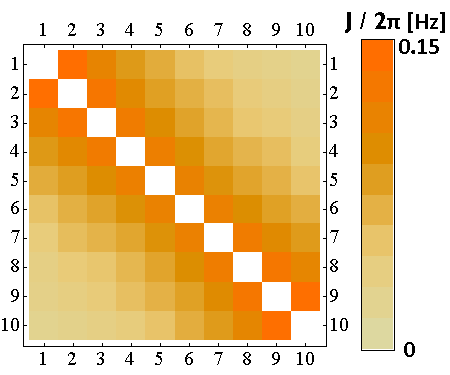}
\end{minipage}
\begin{minipage}[hb]{0.4\textwidth}
\centering
\caption{Coupling constants $J_{n,m}$ in a ten ion crystal at $\omega_{z} = 2\pi\cdot310~\text{kHz}$, for a b=1~T/m magnetic field gradient in the z-direction, thus aligned along the ion crystal. A maximum J$_{n,m}$ of $2\pi\cdot0.15$~Hz/T is achieved.}
\label{jnmaxial}
\end{minipage}
\end{figure}
\vspace{-1cm}
\subsection{Magnetic gradient along the trap axis}
A magnetic field along the trap axis $\vec B =(b_0 + z\cdot b)\vec{e_z}$ results in a field b$_0$, which provides a quantization axis and which splits the $|m=+1/2\rangle$ and the $|m=-1/2\rangle$ spin states by an energy larger than the vibration quanta energy. The second term $z\cdot b$ results in a position dependent Zeeman shift, leading to forces against the axial vibrational modes only which are determined by the block A in the Hesse matrix M yielding:
\begin{equation}
J_{n,m}^{ax} = \frac{g_J^2\mu_B^2}{2\hbar}\frac{b^2}{m\omega_z^2}(A^{-1})_{n,m}
\label{eq_jnm}
\end{equation}
with Bohr magneton $\mu_B$ and Land\'e factor $g_J = 2$. The relative magnitude between different entries of $J$ depends only on the number of ions, whereas the absolute height is proportional to $b^2/\omega_z^2$. Fig.~\ref{jnmaxial} shows the coupling constants for ten ions, normalized to a unit magnetic field gradient of 1 T/m . All couplings are positive and strongest for nearest neighbors with a strength of $2\pi\cdot7.4~\text{Hz}$ for $7~\text{T/m}$ and an axial frequency of $\omega_{z} = 2\pi\cdot310~\text{kHz}$.

\subsection{Magnetic gradient transverse to the trap axis}
\begin{figure}[]
\begin{minipage}[hb]{0.6\textwidth}
\centering
\includegraphics[width = 0.865\textwidth]{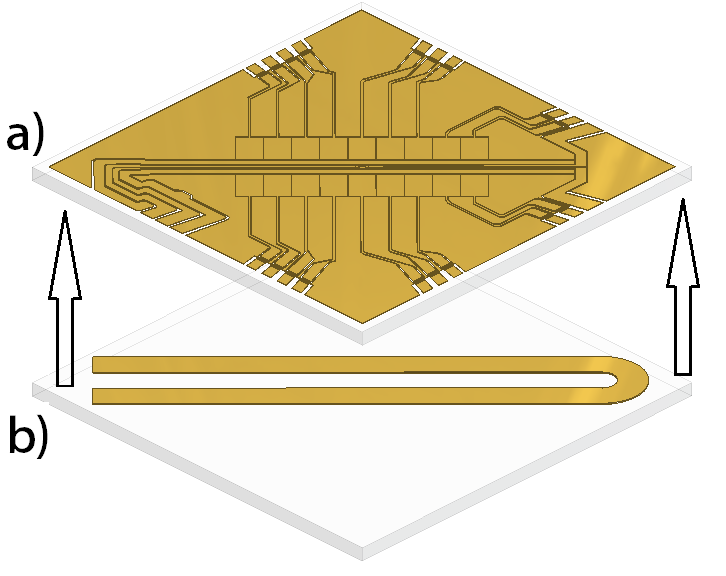}
\end{minipage}
\begin{minipage}[hb]{0.5\textwidth}
\centering
\caption{Sketch of a planar micro ion trap for applying radial magnetic forces: a) Trap chip for ac and dc electric potentials, b) underlying second chip to be mounted directly underneath (a) for magnetic field gradients in the radial direction. Arrows indicate how (b) is mounted under (a).}
\label{H-coiltrap}
\end{minipage}
\end{figure}

We now investigate the interaction when the gradient is aligned along the x-axis and the spin dependent force acts against the transversal eigenmodes which are governed by the anisotropy parameter. The transverse confinement and the lowest radial eigenfrequency can be reduced without modifying the distances between the ions. The Stern-Gerlach force acts by moving the ions transversally away from the trap axis, and the lowest frequency transversal eigenmode, the zig-zag mode governs the interaction. This mode is excited most efficiently with an alternating, anti-ferromagnetic spin alignment. The transversal magnetic gradient couplings $J_{n,m}$ are determined by the matrix B$^x$ (see Appendix~\ref{Hesse}) and shown in Fig.~\ref{jrad} for different values of $\alpha$. When $\alpha$ approaches its critical value, we find coupling strengths of about $2\pi\cdot 158$~kHz, where the frequency of the zig-zag mode $\nu_{zz}$ is at 11~kHz and the transversal field gradient is 23~T/m. Whereas for values far from the zig-zag instability the pattern displays mostly next neighbor negative couplings, the interactions increase when approaching the instability and show a fully antiferromagnetic behavior. The strength of mutual interactions is stronger in the middle of the ion crystal, in contrast to the behavior of the $J_{m,n}$ in the case of axial magnetic gradients in Fig.~\ref{jnmaxial}. The radial restoring forces in an ion crystal are weakest in the middle, while the axial restoring force is weakest at the ends. We  stress that the coupling strengths with transversal magnetic gradients are about three orders of magnitude higher than in the case of interactions driven by a magnetic field gradient of the same magnitude but in the $z$ direction, when lowest radial eigenmode frequency is reduced to about 10~kHz.
\begin{figure}
\centering
\resizebox{1\columnwidth}{!}{
\includegraphics{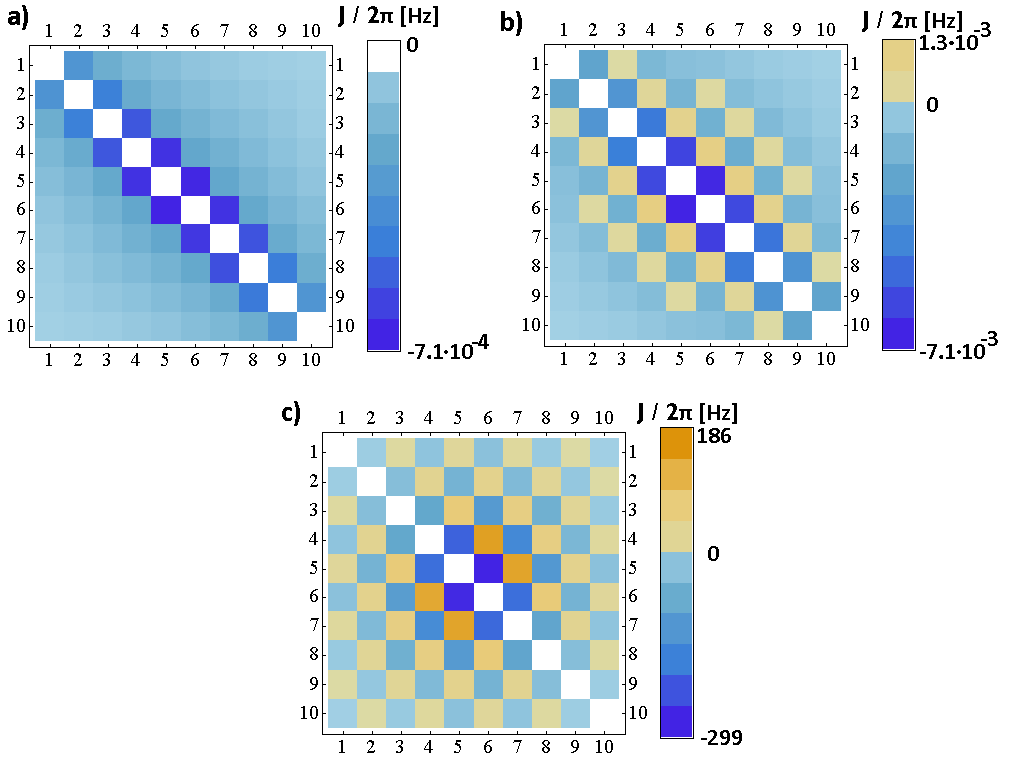} }
\caption{Coupling constant of ten ions normalized to b=1 T/m magnetic field gradient  in the radial direction when approaching the critical anisotropy for ten ions ($\alpha_{crit} = 0.047348$). The three subplots indicate in color coding the transition into a ferromagnetic coupling with red  for positive and  blue for negative values of J.  a) Choosing a value of $\alpha=0.0097819$ results in $\nu_{zz}$=2.8~MHz and a maximum J$_{n,m}$ of $2\pi\cdot7.1~10^{-4}$Hz/T. b) For $\alpha=0.032669$, $\nu_{zz} =1.7$ MHz and J$_{n,m}$ increases to $2\pi\cdot7.1~10^{-3}$Hz/T. c) With $\alpha =0.047347$ we find $\nu_{zz}=$10.7~kHz and J$_{n,m}$ of $2\pi\cdot$300~Hz/T.}
\label{jrad}
\end{figure}
Gradients in the radial direction are generated by the current running in the central wire W$_{1}$ (see Fig. \ref{gradBall}). An alternative scheme for realizing a transversal magnetic field is sketched in Fig.~\ref{H-coiltrap}. A second chip beneath the trap containing just two broad wires with counter propagating currents (narrow U-shaped structure), creates a vertical magnetic field. Working with the same current density as before will give rise to a gradient of about $40~\text{T/m}$ at a height of $400~\mu$m over the plane of the U, fitting perfectly to the ions location. The axial component vanishes, as the bent part of the current loop is far from the ion position.  With the above values we calculate a coupling strength of up to $2\pi\cdot479$~kHz for $\nu_{zz}=11$~kHz.  One question to be answered by future experiments is how accurately the trap control parameters can be adjusted in order to approach the instability point even further for increasing the couplings even into the MHz regime. The MAGIC interaction does not strictly require cooling to the vibrational ground state. Still, at small values of $\nu_{zz}$ the mean phonon number at the Doppler temperature will be excessively high.  One possible way to reduce the phonon number in the mode, that mostly contributes to spin-spin coupling, would be to cool that mode still at high $\nu_{zz}$ and then adiabatically ramp up the value of $\alpha$ for achieving the desired spin-spin coupling strengths.

\subsection{Initialization and read-out of ion spins \label{inizialize}}
Before the application of the spin-dependent force, one might like to align all spins in one direction. This can be done by standard optical pumping e.g. with $\sigma +$ laser light such that the m=+1/2$\equiv |\uparrow\rangle$ state is occupied. An even higher fidelity of the initialization employs frequency-resolved optical pumping yielding a transfer of 99.6\% \cite{POSCH}. A subsequent optical Raman pulse on the entire ion crystal with a well chosen length of $\theta = \pi/2$ generates the superposition state $(|\uparrow\rangle + |\downarrow\rangle)/\sqrt{2}$ for each ion. The read-out would be performed by a detection of state-dependent fluorescence, a well established tool in experimental quantum information with trapped ions. The advantage is that, with careful application, the read-out fidelity can reach values exceeding 99.99\% \cite{MYER08}, and that this technique also resolves the individual ion sites such that the spin order after the interaction can be directly observed by state-dependent fluorescence on  CCD-camera images.

\subsection{Selective excitations via resonant current \label{AC_Current}}
So far, we have discussed spin-spin interactions when a static magnetic field gradient is applied. The small induction of the wire loop that generates the gradient allows one to apply alternating currents $I(t)=I_0 \sin(\omega t)$, with frequencies of a few kHz up to even a few MHz and where $\omega$ may be tuned close to one of the motional frequencies $2\pi\nu_m$ in the ion crystal. After a proper initialization, this will predominantly excite the ions at the positions corresponding with the eigenvectors of that specific mode $m$, thus selecting spins at specific positions of the crystal \cite{NAEGERL98,DREWSEN}. If for example, the breathing mode in a 3-ion crystal at $(2\pi) \sqrt{3} \nu_{z}$ is excited by an alternating gradient in the z-direction, only the outermost ions couple, and the middle ion is excluded from the interaction. When multiple frequencies are applied on ion crystals with $I(t)=\sum_j I_j \sin(\omega_j t)$, one may excite superpositions of eigenmodes, where arbitrary ion excitations can be achieved.

The second advantage of resonantly applied gradients results from a resonant enhancement, such that the interaction strengths $J_{n,m}$ are increased as compared to their static values. We propose the generation of Schr\"odinger  cat states \cite{POSCH2010}  $|\Psi_f\rangle=\frac{1}{\sqrt{2}}(|\uparrow,\alpha(t)\rangle+i|\downarrow,-\alpha(t)\rangle)$, where the ion crystal is first initialized into the superposition spin state and, via the magnetic gradient forces, the motional degree of freedom is entangled with the spin state of an ion when a single ion is excited very close to the axial vibration mode with a detuning $\delta = \omega - 2\pi\nu_{z}$. Large values of displacement $\alpha$ are reached for small $\delta$. In analogy to the application of spin dependent light forces a two-ion crystal gate operation similar to the geometric gate operation with state dependent ac light forces \cite{MOEL99,Leib2003,ROOS2008,KIM2009,EDWARDS2010} can be realized \cite{ospelkaus11}, where a transient motional excitation is generated only for specific spin (qubit) states, and results in a conditional phase shift of the wave function.

It should be noted that the investigation of the linear to zig-zag configuration in a linear string \cite{Retzker08,Shimshoni11} is a quantum phase transition which has attracted much attention recently. However, the experimental realization has not yet been achieved. We state that the transverse magnetic field gradients would be ideally suited to drive the system in a spin-dependent way from one phase  to the other \cite{ivanov}.

\begin{figure}[thb]
\begin{minipage}[hb]{0.5\textwidth}
\centering
\includegraphics[width = 0.865\textwidth]{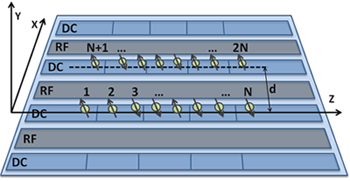}
\end{minipage}
\begin{minipage}[hb]{0.5\textwidth}
\centering
\caption{Sketch of a planar trap for trapping multiple parallel ion strings.}
\label{multiString}
\end{minipage}
\end{figure}

\section{Spin-Spin interactions in two dimensional ion crystals \label{2D}}
Having discussed the tunability of spin interactions with linear ion crystals, we aim for simulations of larger systems of interacting spins and more complex geometries, that are only possible in two dimensional systems. In the following, we will outline in detail, how planar ion traps with integrated magnetic field gradients can reach such advanced goals of quantum simulation. For this we present a calculation of vibrational modes for 2D-ion crystals, followed by the results of spin-spin interactions in such system. Our versatile, dynamic and even possibly fast control of the ion crystals positions in the two-dimensional lattice, can be seen in contrast to proposals for static planar arrays \cite{Chiaverini08,Schmied09}. It appears that the advantage of our scheme is that it may allow for investigating frustrated spin systems with tunable interactions. The geometry of the latices can be changed on the run. This may be also reached by a dynamical control of the radio frequency, as proposed recently \cite{kumph}. Furthermore, our scheme gives a clear route how the ions crystals are loaded in a remote, unproblematic since wider part of the trap and then being transported to the narrow trap interaction region for the two dimensional spin-spin interactions. As the loading of linear shaped planar traps has been successful in many labs including ours, this scheme appears to be in reach of the experimental possibilities.

A sketch of a planar trap for the storage of two linear ion strings parallel to each other is shown in Fig.~\ref{multiString}, extending the generic planar five-wire trap design only slightly, here by one rf and one dc electrode.  Lithographically produced micro structures \cite{BROWN2011} will allow for trapping multiple strings. For describing the interactions, the potential in eq.~\ref{eqn_pot} has to be slightly modified by adding the electrostatic coupling between ions which sit in different strings,
\begin{equation*}
\label{eqn_pot2}
V = m\omega_z^2\sum_{n=N+1}^{2N}\Big(\sum_{i=1}^2\frac{1}{2\alpha_i}(x_{n,i})^2+\frac{1}{2\alpha_x}(x_{n,0}-d)^2\Big),
\end{equation*}
where $N$ now denotes the number of ions in one chain and $d$ the distance between the trap centers in the $x$-direction. The upper limit of the sums in the second part of  eq.~\ref{eqn_pot} therefore runs to $2N$. The general form of the Hessian M changes to dimension $2N$, and M is still blocked, but no longer diagonal, for there are non-zero entries coupling radial and axial modes. Only in the limit of $d$ to infinity does M simplify to a diagonal matrix with separated entries $(B^x,B^x,B^y,B^y,A,A)$ describing two independently oscillating linear strings.

\subsection{Eigenmodes of  parallel arranged linear ion crystals}
\begin{figure}
\begin{minipage}[hb]{0.35\textwidth}
\centering
\caption{Eigenfrequencies $\nu$ of two ion strings with two ions each confined in a 2-dimensional array, plotted in multiples of $\omega_z/(2\pi)$ as a function of the distance $d$ between the axes of the strings for $N=2$. The values correspond to a radial confinement with $\omega_{x,y,z}$ of $3.42~\text{MHz}$, $2.72~\text{MHz}$, and $1.04~\text{MHz}$, when $d$ approaches infinity. The plot shows the dependence of both radial and both rocking modes (green), the breathing (red) and the axial center-of-mass mode (green), from top to bottom. The anti-modes are plotted in bright color and the co-modes in dark, respectively. The ion's motion in one string is indicated by the direction of the arrows for large distances.}
\label{eigenvalues}
\end{minipage}
\begin{minipage}[hb]{0.65\textwidth}
\centering
\includegraphics[width = 0.865\textwidth]{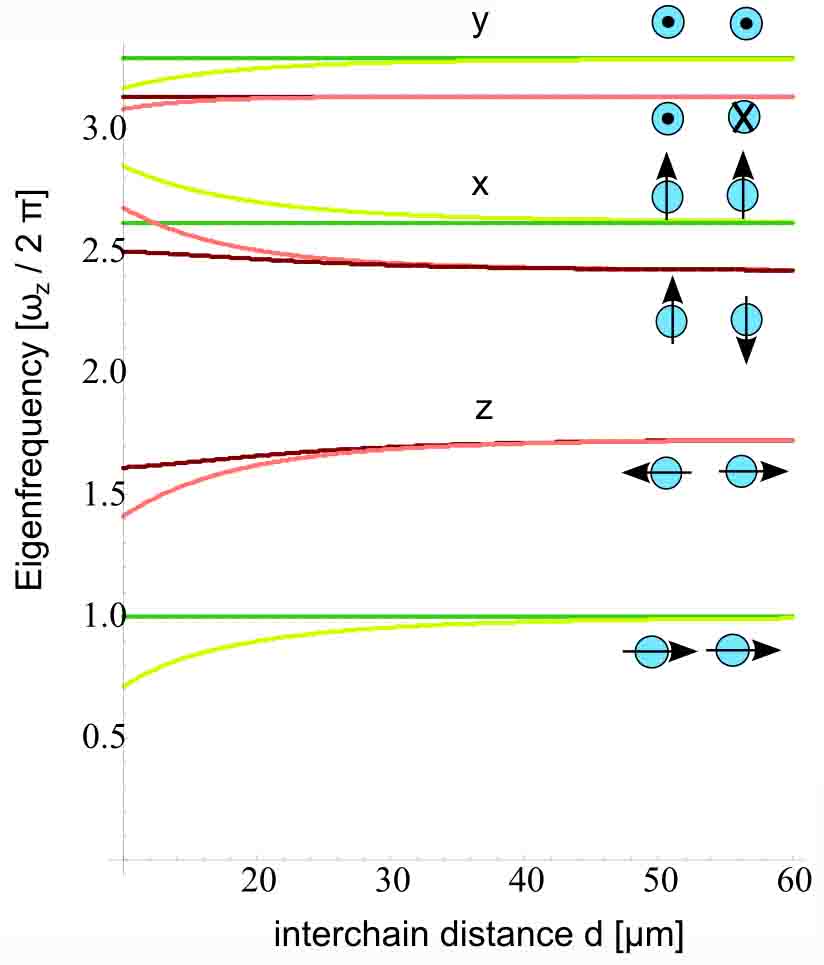}

\end{minipage}
\end{figure}

In a two-dimensional ion crystal, any movement of one chain will influence the other such that the separation in radial and axial eigenmodes fails, and 3$\times$(2$N$) eigenmodes exist. In the following we exemplify the effects with a system of two strings each containing two ions. Ions in one string can vibrate in common or contrariwise, commonly denoted as center-of-mass and breathing mode respectively. The respective frequencies $\nu$ are characterized by the square root of the Hessian matrix's eigenvalues given in multiples of $m \omega_z^2$, namely $1$ and $\sqrt{3}$ in the axial direction. For the radial direction mode frequencies depend on the value of $\alpha$, and are denoted as the radial center-of-mass and the rocking mode. In combination with the second string arranged parallel to the first one, we find that the four modes of one string split up in a co- and an anti-mode, plotted over the distance $d$ in Fig.~\ref{eigenvalues} (tables of eigenmodes in the app.~\ref{tab_trap}). The frequency shifts are stronger if the chains are moving in the anti-configurations. The modes with dominant amplitude along the $x$-axis in the direction of $d$ display increasing eigenvalues, whereas the $z$ dominated transversal modes are decreasing in frequency. With many ions per chain and below a certain $\alpha$ both chains will undergo, similar to the case of a single string, a transition to a  zig-zag-zig-zag configuration. However at small values of $d$ both chains generate a stabilizing effect onto each other due to the mutual Coulomb repulsion, thus remaining in a linear configuration well beyond the critical value of $\alpha$ for a single ion crystal of same size. We also calculated the resulting equilibrium positions for the two-dimensional crystal and observe a small bending of the strings due to a comparatively greater shifting of the ions in the middle of the strings. In the case of ten ions at a distance of $d = 50~\mu$m with $\alpha_x =0.00975$, the numerical solution yields a relative increase of the distance between the strings of 1.4~10$^{-3}$ and a relative increase of the inter-ion distances in one string of 1.4~10$^{-2}$.  This effect might be counteracted by proper compensation.
\begin{figure}
\begin{center}
\resizebox{0.85\columnwidth}{!}{
\includegraphics{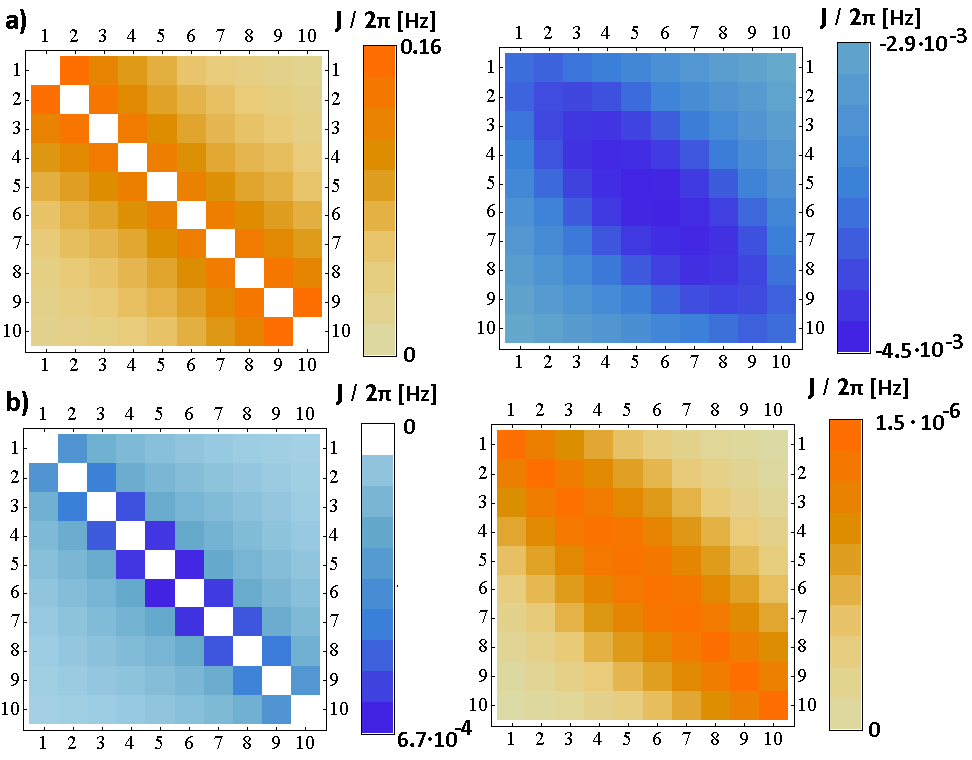} }
\resizebox{0.85\columnwidth}{!}{
\includegraphics{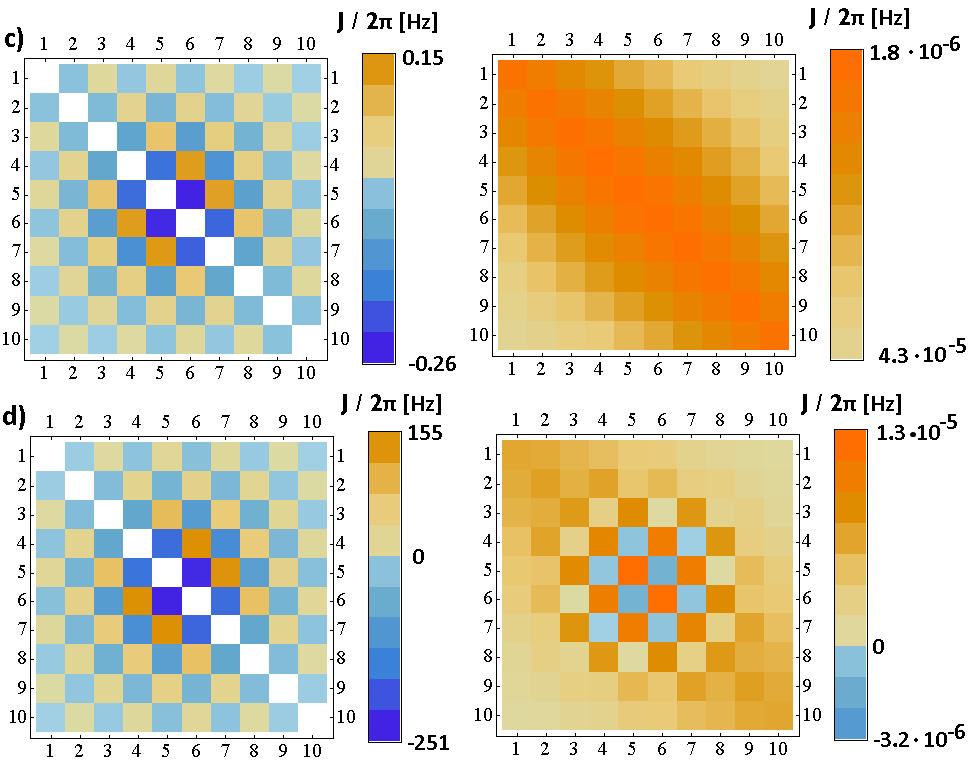} }
\end{center}
\caption{Coupling strength for two chains with ten ions at a distance of $50~\mu\text{m}$ normalized to $\text{b}=1~\text{T/m}$ unit magnetic gradient. The inter-string coupling blocks (off diagonal elements) are found in a 3D plot below the corresponding matrix plot. The gradient magnetic field is applied either in the z (axial) direction in case of subplot (a) or in x (transversal) for the subplots (b,c,d). In (a) the anisotropy parameter is fixed well above the instability to the zig-zag transition at $\alpha = 0.009782$ with $\omega_z=310$~kHz yielding ferromagnetic coupling within the strings and much weaker anti-ferromagnetic interactions between both strings. For a transverse magnetic gradient, the anisotropy parameter is varied from $\alpha =0.009782$, $\nu_{zz} = 2.8$~MHz (in b) to $\alpha = 0.047347$, a value which corresponds to the instability transition for a single linear string, at $\nu_{zz}$=356~kHz (in c) and $\alpha =0.048014$ with $\nu_{zz} = 11.5$~kHz (in d).}
\label{2ketten}
\end{figure}

\begin{figure}[b]
\begin{minipage}[hb]{0.5\textwidth}
\centering
\includegraphics[width = 0.865\textwidth]{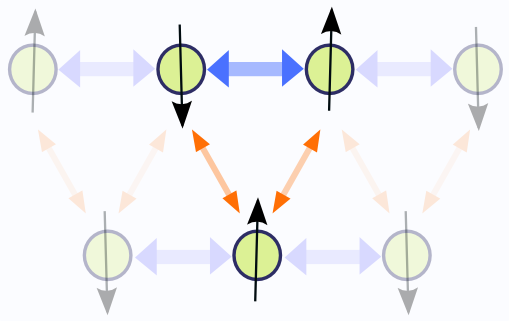}
\end{minipage}
\begin{minipage}[hb]{0.5\textwidth}
\centering
\caption{Displacement of two linear ion strings with respect to each other creates a triangular geometry. The designed coupling of spins could be used to create frustrated ion crystals.}
\label{frust}
\end{minipage}
\end{figure}

\subsection{Magnetic coupling with axial and radial gradient fields} \label{ferroantiferro}
Surprisingly complex magnetic interactions arise from the application of magnetic field gradients. In Fig.~\ref{2ketten} the interaction strengths  are shown for two chains with ten ions with a normalized 1 T/m  magnetic field gradient applied either purely in $z$ or purely in  the $x$ direction. The coupling matrix is separated into four blocks, where the diagonal ones represent the intra chain coupling, as discussed in the previous sections. The off-diagonal blocks describe the interaction between chains.

In the case of a field gradient aligned along the direction of the ion strings (in Fig.~\ref{2ketten}a), the ferromagnetic coupling within a string is recovered. A very weak coupling to the neighbor string is anti-ferromagnetic, and thus the lowest energy state of the system will lead to two spin chains, where all spins within a string are parallel, but oppositely oriented with respect to the other string. For $\text{b}=40~\text{T/m}$ the maximum magnitude of J is $2\pi\cdot$260~Hz within one string and 2$\pi\cdot$7.2~Hz between strings. Even more interesting behavior is observed in the cases displayed in Fig.~\ref{2ketten}~b,c,d for the transverse gradient. When approaching the zig-zag instability, the coupling within the strings develops an anti-ferromagnetic order, while the coupling between different strings undergoes a transition from ferromagnetic to anti-ferromagnetic order.  Far  above the critical value of $\alpha$, at $\alpha =0.009782$ (in Fig.~\ref{2ketten}b) we calculate very small values of J with 2$\pi\cdot$1.1~Hz and 2$\pi\cdot$ 2.4$\cdot10^{-3}$~Hz. Approaching closer to the instability we find 2$\pi\cdot$416~Hz and 2$\pi\cdot2.9\cdot10^{-3}$~Hz (in c) and when the zig-zag mode is tuned down to $\nu_{zz}=11.5$~kHz the couplings increase to 2$\pi\cdot$402~kHz and 2$\pi\cdot2.1\cdot10^{-2}$~Hz. Even though the off-diagonal elements of the coupling matrix are still rather weak as compared with the diagonal elements, we expect that the technique of alternating near-resonant  field gradients as discussed in Sec.~\ref{AC_Current} will enhance these elements such that an inter-string coupling is possible to observe. Evidently, bringing the two ion strings to a distance even closer than $d=50~\mu$m would also lead to significantly increased off-diagonal couplings J$_{n,m}$.

\subsection{Frustration effects}
A configuration of two strings parallel to each other may also be used to investigate frustration effects in a spin system. For this we propose shifting one of the chains along the z axis by half the inter-ion distance in the string. Now, ions of one string are at equal distance to \emph{two} ions in the other chain. With a ferromagnetic interaction between both strings, and an anti-ferromagnetic interaction within the strings, the ions minimize energy when aligning with two counterparts in the other string, but those have different spins, as shown in Fig. \ref{frust}.  If  the bending of the strings can't be overcome, this would be a major drawback to this scheme, since in that case only one triple of ions can be exactly frustrated at one time. The calculation of the resulting spin orientation is beyond the scope of this paper, but in principle it is possible to adjust the intra- and inter-string coupling strengths. As an example, by bringing the strings as close as $20~\mu$m together and with a relatively loose axial confinement of $\omega_z = 2\pi\cdot 155$~kHz and $\alpha=0.0024$ the intra-chain interaction strength drops to $2\pi\cdot1.02\cdot 10^{-4}$~ Hz, whereas the inter-chain coupling strengths are about $2\pi\cdot2.83\cdot 10^{-5}$~Hz, thus leading to a tunable ratio, in this case of about 3.6. In a future experiment, one may control the mutual interactions by a variation of $\alpha$ and the magnetic field gradient, also by positioning of the strings with small changes of the trap control voltages such that it appears well within reach to switch back and forth between rectangular lattice configuration and a triangular one resulting in frustration effects. The ion model system might allow one to contribute to the vivid theoretical discussion of quasi one dimensional structures in interacting quantum wires \cite{Meyer}, where the quantum conductance depends on the filling factor, as when increasing the number of electrons in a quantum wire, they undergo a transition to a zig-zag configuration. It may be further investigated how trapped ion crystals could be employed to simulate transport in quasi-linear solid state structures. One might use trapped ion strings which are moved along each other as a model system with versatile and tunable properties, as a model of genuine solid state physics problems.

\section{Conclusion}
We have outlined a scheme for investigating spin-spin interaction with the help of planar ion traps. We obtained a description of the magnetic interaction properties for a one or two dimensional crystal, strongly depending on both the direction and the strength of the magnetic gradient that is applied. Ferromagnetic or anti-ferromagnetic configurations are generated, and with a control of the gradient by two independent current-carrying wires as sources for the fields it is possible to direct it in any direction of the x-z plane. The strength of the interaction is proportional to two easily accessible parameters, namely the magnetic gradient via the chosen current and the axial trap frequency determined by the applied dc voltages. Although the electrode geometry of a planar trap is fixed, it still offers a range of possible trapping frequencies to realize potential wells that confine the ions very tightly or allowing the crystal to restructure itself e.g. into the zig-zag configuration. The segmented outer dc electrodes can be used to displace ion strings against each other or to split up one chain. Offsets on the rf electrodes could be used to modify the distance between chains, thus one can shift the eigenfrequencies of the system and manipulate the inter chain coupling. A rich plethora of interactions is possible in such two-dimensional quantum spin crystals and the related physics opens up a wide field of multi-particle spin simulations. Perhaps among the most striking features is the possibility to simulate frustrated spins in two-dimensions. In the paper we have discussed all these future possibilities in very close correspondence to the existing experimental parameters, thus in a fully realistic setting.

The authors would like to thank U. Poschinger, M. Plenio, and A. Retzker for helpful discussions. CW acknowledge support by the EU commission within the STREP-PICC. FSK and RF acknowledge financial support by the German-Israeli Science Foundation (GIF), FSK the German Science Foundation within the SFB-TRR21, the European Commission within the IP-AQUTE and by the Office of the Director of National Intelligence (ODNI), Intelligence Advanced Research Projects Activity (IARPA), through the Army Research Office grant W911NF-10-1-0284.  All statements of fact, opinion or conclusions contained herein are those of the authors and should not be construed as representing the official views or policies of IARPA, the ODNI, or the U.S. Government. We acknowledge the clean room support by P. Ziemann and A. Plettl.

\appendix

\section{Calculation of the Hesse Matrix \label{Hesse}}

Since we have $N$ ions and in general three spatial degrees of freedom the Hessian of Eq. \ref{eqn_pot} can be written  formally as
\begin{eqnarray*}
		 M_{jN+o,kN+p}&=&m\omega_z^2\Bigg(\frac{\delta_{jk}\delta_{op}}{\alpha_j }+\frac{l^3}{2}\sum_{n\neq m}(\delta_{on}-\delta_{om})(\delta_{np}-\delta_{mp})\cdot\\
					&&\cdot\bigg[\frac{3(x_{n,k}-x_{m,k})(x_{n,j}-x_{m,j})}{|\mathbf x_n - \mathbf x_m|^5}-\frac{\delta_{kj}}{|\mathbf x_n - \mathbf x_m|^3}\bigg]_{\mathbf{\bar x}_{n},\mathbf{\bar x}_{m}}\Bigg)\label{eqn_M},\end{eqnarray*} with $o$ and $p$ denoting the  ion's number, $j$ and $k$ the spatial direction and $\delta_{jk}$ the Kronecker delta.
For a single string and as long as one stays below the critical value of $\alpha$, ions align linearly and thus $M$ reduces to a block diagonal matrix \cite{Enzer00,James98} with entries $m\omega_z^2 (B^x,B^y,A)$  where \begin{eqnarray*}
A_{n,m}& =&  \begin{cases} 1+2\sum_{\substack{p=1\\p\neq m}}^{N}\frac{l^3}{|\bar z_{m}- \bar z_{p}|^3}&\text{ if } n=m\\
\frac{-2l^3}{|\bar z_{m}- \bar z_{n}|^3}&\text{ if } n\neq m
\end{cases}\\
B^i_{n,m}&=&\Big(\frac{1}{\alpha_i}+\frac{1}{2}\Big)\delta_{nm}-\frac{1}{2}A_{nm}.
\end{eqnarray*}

In the case of two strings with each holding $N$ ions, the two sums in $M$ can be split apart in intra string and inter string components:
\begin{eqnarray*}
\sum\limits_{n=1}^{2N}\sum\limits_{\substack{m=1\\ m\neq n}}^{2N}= \sum\limits_{\substack{n,m=1\\ m\neq n}}^{N} +\sum\limits_{\substack{n,m=N+1\\ m\neq n}}^{2N}+2\sum\limits_{n=1}^N\sum\limits_{m=N+1}^{2N}= (a)+(b)+(c).
\end{eqnarray*}
Sum ($a$) will yield to the matrices $A$ and $B$ for one chain,  ($b$) to that of the other chain, such that one has a matrix $(B^{x1},B^{x2},B^{y1},B^{y2},A^1,A^2)$.
Part ($c$) however leads to new entries in $M$, which reads with $\delta^{op}_{nm} = (\delta_{no}-\delta_{mo})(\delta_{np}-\delta_{mp})$:
\begin{eqnarray*}
C^{xx}_{op}&=& l^3\sum_{n,m}\delta^{op}_{nm}\frac{2 (x_n-x_m)^2-(z_n-z_m)^2}{|\mathbf x_n - \mathbf x_m|^5}\\
C^{zz}_{op} &=& l^3\sum_{n,m}\delta^{op}_{nm}\frac{2(z_n-z_m)^2-(x_n-x_m)^2}{|\mathbf x_n - \mathbf x_m|^5}\\
C^{xz}_{op} &=& l^3\sum_{n,m}\delta^{op}_{nm}\frac{3(x_m-x_n)(z_n-z_m)}{|\mathbf x_n - \mathbf x_m|^{5}} = C^{zx}_{op}\\
C^{yy}_{op} &=& l^3\sum_{n,m}\delta^{op}_{nm}\frac{-1}{|\mathbf x_n - \mathbf x_m|^3}\\
C^{xy}_{op} &=& C^{yx}_{op}= C^{zy}_{op}= C^{yz}_{op} = 0\\
\end{eqnarray*}

\section{Trapping parameters and eigenmodes}
Table (\ref{tab_trap}) shows possible values for the voltages applied to three neighboring  pairs of dc electrodes to achieve axial confinement. The electrode pair in the middle is held at voltages $U_2$ and both outer pairs at $U_1$.
\begin{table}[h]{\footnotesize
\centering
\caption{Influence of the axial confinement on radial frequencies and resulting anisotropy. }
\label{tab_trap}       
\begin{tabular}{lllllll}
\hline\noalign{\smallskip}
$U_1$[V] & $U_2$[V] & $\omega_x/2\pi$[MHz] & $\omega_y/2\pi$[MHz] & $\omega_z/2\pi$[MHz] & $\alpha_x$ & $\alpha_y $\\
\noalign{\smallskip}\hline\noalign{\smallskip}
1.30&-1.22&3.14&3.20&0.31&0.00975&0.00939\\
4.12&-3.87&3.06&3.25&0.55&0.0325&0.0288\\
5.75&-5.40&3.01&3.28&0.65&0.0470&0.0396\\
5.82&-5.46&3.00&3.28&0.66&0.0476&0.0400\\
6.53&-6.13&2.98&3.29&0.69&0.0542&0.0446\\
13.0&-12.2&2.77&3.39&0.98&0.125&0.0835\\
23.2&-21.7&2.41&3.55&1.30&0.297&0.135\\
\noalign{\smallskip}\hline
\end{tabular}}
\end{table}

\begin{table}{\tiny
\caption{Eigenmodes in a distance $d=100~\mu$m. }
\label{tab_eigen}       
\begin{tabular}{llllllllllll}
\hline\noalign{\smallskip}
$x_1$&$ x_2$& $x_3$& $x_4$&$y_1$& $y_2$& $y_3$&$y_4$&$z_1$&$z_2$&$z_3$& $z_4$\\
\noalign{\smallskip}\hline\noalign{\smallskip}
0& 0& 0& 0& -1/2& -1/2& -1/2& -1/2& 0& 0& 0& 0\\0& 0& 0&
  0& -1/2& -1/2& 1/2& 1/2& 0& 0& 0& 0\\0& 0& 0& 0& -1/2& 1/2& -1/2&
  1/2& 0& 0& 0& 0\\0& 0& 0& 0& -1/2& 1/2& 1/2& -1/2& 0& 0& 0&
  0\\-1/2& -1/2& 1/2& 1/2& 0& 0& 0& 0& 0& 0& 0& 0\\1/2& 1/2& 1/2&
  1/2& 0& 0& 0& 0& 0& 0& 0& 0\\-1/2& 1/2& 1/2& -1/2& 0& 0& 0& 0& 0&
  0& 0& 0\\1/2& -1/2& 1/2& -1/2& 0& 0& 0& 0& 0& 0& 0& 0\\0& 0& 0&
  0& 0& 0& 0& 0& 1/2& -1/2& 1/2& -1/2\\0& 0& 0& 0& 0& 0& 0& 0& 1/
  2& -1/2& -1/2& 1/2\\0& 0& 0& 0& 0& 0& 0&
  0& -1/2& -1/2& -1/2& -1/2\\0& 0& 0& 0& 0& 0& 0& 0& -1/2& -1/2& 1/
  2& 1/2\\
\noalign{\smallskip}\hline
\end{tabular}
\caption{Eigenmodes in a distance $d=50~\mu$m.  }
\label{tab_eigen}       
\begin{tabular}{llllllllllll}
\hline\noalign{\smallskip}
$x_1$&$ x_2$& $x_3$& $x_4$&$y_1$& $y_2$& $y_3$&$y_4$&$z_1$&$z_2$&$z_3$& $z_4$\\
\noalign{\smallskip}\hline\noalign{\smallskip}
0& 0& 0& 0& -1/2& -1/2& -1/2& -1/2& 0& 0& 0& 0\\0& 0& 0&
  0& -1/2& -1/2& 1/2& 1/2& 0& 0& 0& 0\\0& 0& 0& 0& 1/2& -1/2& 1/
  2& -1/2& 0& 0& 0& 0\\0& 0& 0& 0& 1/2& -1/2& -1/2& 1/2& 0& 0& 0&
  0\\-1/2& -1/2& 1/2& 1/2& 0& 0& 0& 0& -5E-5& 5E-5& -5E-5&
  5E-5\\1/2& 1/2& 1/2& 1/2& 0& 0& 0& 0& 0& 0& 0& 0\\-1/2& 1/2&
  1/2& -1/2& 0& 0& 0& 0& 0& 0& 0& 0\\1/2& -1/2& 1/2& -1/2& 0& 0& 0&
  0& 5E-5& 5E-5& -5E-5& -5E-5\\-5E-5& -5E-5&  -5E-5& 5E-5& 0& 0& 0& 0& 1/2& -1/2& 1/2& -1/2\\0& 0& 0& 0& 0&
  0& 0& 0& 1/2& -1/2& -1/2& 1/2\\0& 0& 0& 0& 0& 0& 0&
  0& -1/2& -1/2& -1/2& -1/2\\5E-5& -5E-5& 5E-5& -5E-5&
  0& 0& 0& 0& -1/2& -1/2& 1/2& 1/2\\
\noalign{\smallskip}\hline
\end{tabular}
\caption{Eigenmodes in a distance $d=20~\mu$m.  }
\label{tab_eigen}       
\begin{tabular}{llllllllllll}
\hline\noalign{\smallskip}
$x_1$&$ x_2$& $x_3$& $x_4$&$y_1$& $y_2$& $y_3$&$y_4$&$z_1$&$z_2$&$z_3$& $z_4$\\
\noalign{\smallskip}\hline\noalign{\smallskip}

 0& 0& 0& 0& 1/2& 1/2& 1/2& 1/2& 0& 0& 0& 0\\
0& 0& 0& 0& -1/2& -1/2& 1/2& 1/2& 0& 0& 0& 0\\
0& 0& 0& 0& 1/2& -1/2& 1/2& -1/2& 0& 0& 0& 0\\
0& 0& 0& 0& -1/2& 1/2& 1/2& -1/2& 0& 0& 0& 0\\
-1/2& -1/2& 1/2& 1/2& 0& 0& 0& 0& -1.65E-3& 1.65E-3& -1.65E-3&
  1.65E-3\\
1/2& -1/2& -1/2& 1/2& 0& 0& 0& 0& 0& 0& 0& 0\\
-1/2& -1/2& -1/2& -1/2& 0& 0& 0& 0& 0& 0& 0& 0\\
-1/2& 1/2& -1/2& 1/2& 0& 0& 0& 0& -1.64E-3& -1.64E-3& 1.64E-3&
  1.64E-3\\
1.65E-3& 1.65E-3& -1.65E-3& -1.65E-3& 0& 0& 0& 0& -1/2& 1/
  2& -1/2& 1/2\\0& 0& 0& 0& 0& 0& 0& 0& 1/2& -1/2& -1/2& 1/2\\
0& 0& 0& 0& 0& 0& 0& 0& -1/2& -1/2& -1/2& -1/2\\
-1.64E-3& 1.64E-3& -1.64E-3& 1.64E-3& 0& 0& 0& 0& 1/2& 1/
  2& -1/2& -1/2\\
\noalign{\smallskip}\hline
\end{tabular}
}\end{table}

The following tables contain the eigenmodes of a system with two ion chains, each containing 2 ions.  The subscript refers to the $n$th ion, where ion $1$ and $2$ are in one chain and ion $3$ and $4$ are in the second chain. The different modes are sorted  in descending order according to the eigenvalues.


\begin{thebibliography}{plain}
\bibitem{Feynman82}
R. Feynman, Int. J. Theor. Phys. \textbf{21}  (1982) 467.

\bibitem{Barreiro10}
J. T. Barreiro, et. al., Nature Phys. \textbf{6} (2010) 943.

\bibitem{Weimer10}
H. Weimer, M. M\"uller, I. Lesanovsky, P. Zoller, H. P. B\"uchler,
Nature Phys. \textbf{6} (2010) 382.

\bibitem{Jaksch05}
 D. Jaksch, P. Zoller, Ann. Phys. \textbf{52}, (2005) 315.

\bibitem{Cirac00}
J. I. Cirac, P. Zoller, Nature \textbf{404},  (2000) 579.

\bibitem{Cirac95}
J. I. Cirac,  P. Zoller, Phys. Rev. Lett. \textbf{20}  (1995) 4091.

\bibitem{Bloch2008}
I. Bloch, J. Dalibard, W. Zwerger, Rev. Mod. Phys. \textbf{80} (2008) 885.

\bibitem{Sen2008}
A. Sen(De), U. Sen, J. Dziarmaga, A. Sanpera, M. Lewenstein, Phys. Rev. Lett. \textbf{101}   (2008) 187202.

\bibitem{Santos2004} L. Santos, et. al., Phys. Rev. Lett. \textbf{93} (2004) 030601.

\bibitem{Gorshkov10}
A. V. Gorshkov, et al.,  Nature Phys. \textbf{6}   (2010) 289.

\bibitem{Taie10}
S. Taie, et. al., Phys. Rev. Lett. \textbf{105}  (2010) 190401.

\bibitem{Porras04}
D. Porras, J. I. Cirac, Phys. Rev. Lett. \textbf{92}  (2004) 207901.

\bibitem{Schaetz08} A. Friedenauer, H. Schmitz, J. Gl\"uckert, D. Porras and T. Sch\"atz,
Nature Phys. \textbf{4} (2008)  757.

\bibitem{Kim10}
K. Kim et. al., Nature \textbf{465} (2010) 590.

\bibitem{ROOS2010a} J. Casanova, J. J. Garc\'ia-Ripoll, R. Gerritsma, C. F. Roos, E. Solano, Phys. Rev. A \textbf{82}  (2010) 020101(R).

\bibitem{GER2010} R. Gerritsma, G. Kirchmair, F. Z\"ahringer, E. Solano, R. Blatt, C. F. Roos, Nature \textbf{463} (2010) 68.

\bibitem{Hauke10}
P. Hauke, F. Cucchietti, A. M\"uller-Hermes, M. Ba\~nuls,
J. I. Cirac, M. Lewenstein, New J. Phys. \textbf{12} (2010) 113037.

\bibitem{Wunderlich02}
C. Wunderlich, \textit{Conditional spin resonance with trapped ions \emph{in}  Laser Physics at the limit} (Springer, Berlin 2002) 261, arXiv:quant-ph/0111158.

\bibitem{Wunderlich09} M. Johanning, et. al., Phys. Rev. Lett. \textbf{ 102}, 073004 (2009)

\bibitem{WunderlichPRA}
H. Wunderlich, C. Wunderlich, K. Singer, F. Schmidt-Kaler,
Phys. Rev. A \textbf{79} (2009) 052324.

\bibitem{Osp08}
C. Ospelkaus, et. al.,  Phys. Rev. Lett. \textbf{101} (2008) 090502.

\bibitem{Johanning09}
M. Johanning, A. Var\'on, C. Wunderlich, J. Phys. B: At. Mol. Opt. Phys. \textbf{42} (2009) 154009.

\bibitem{Chiaverini08}
J. Chiaverini, W. E. Lybarger, Jr., Phys. Rev. A \textbf{77} (2008) 022324.

\bibitem{BGU} Fabricated at the Ben-Gurion University Nanofabrication Facility (www.bgu.ac.il/nanofabrication)

\bibitem{Chiaverini05}
J. Chiaverini, et. al., Quant. Inf. Comput. \textbf{5} (2005) 419.

\bibitem{Kielpinski02}
D. Kielpinski, C. R. Monroe, D. J. Wineland, Nature \textbf{417}  (2002) 709.

\bibitem{Daniilidis10} N. Daniilidis, et. al., New J. Phys. \textbf{13} (2011) 013032.

\bibitem{Seidelin06}
S. Seidelin, et. al., Phys. Rev. Lett. \textbf{96} (2006)  253003.

\bibitem{Labaziewicz08}
J. Labaziewicz, et. al., Phys. Rev. Lett. \textbf{100}  (2008) 013001.

\bibitem{Wang09} S. X. Wang, J. Labaziewicz, Y. Ge, R. Shewmon, I. L. Chuang, Appl. Phys. Lett. \textbf{94} (2009) 094103.

\bibitem{Amini08}J. M. Amini, J. Britton, D. Leibfried, and D. J. Wineland,  arXiv:0812.3907.

\bibitem{House08}
M. G. House, Phys. Rev. A \textbf{78},  (2008) 033402.

\bibitem{Singer10} K. Singer, et. al., Rev. Mod. Phys. \textbf{82} (2010) 2609.

\bibitem{Huber10} G. Huber, F. Ziesel, U. G. Poschinger, K. Singer, F. Schmidt-Kaler, Appl. Phys. B: Lasers and Optics \textbf{100} (2010) 725.

\bibitem{Hellwig10}
M. Hellwig, A. Bautista-Salvador, G. Werth, F. Schmidt-Kaler, New J. Phys. \textbf{12} (2010) 065019.

\bibitem{radia1}
P. Elleaume, O. Chubar, J. Chavanne, \textit{Computing 3D Magnetic Field from Insertion Devices}, proc. of the PAC97, Conference May (1997) 3509.

\bibitem{radia2}
O. Chubar, P. Elleaume, J. Chavanne, \textit{A 3D Magnetostatics Computer Code for Insertion devices}, SRI97 Conference August 1997, J. Synchrotron Rad. \textbf{5}  (1998) 481.

\bibitem{Groth04}
S. Groth, et. al., Appl. Phys. Lett. \textbf{85}  (2004) 2980.

\bibitem{Zhu06}
S.-L. Zhu, C. Monroe, L.-M. Duan, Phys. Rev. Lett. \textbf{97} (2006) 050505.

\bibitem{James98}
D. F. V. James, Appl. Phys. B \textbf{66} (1998) 181.

\bibitem{Enzer00}
D. G. Enzer, et. al.,  Phys. Rev. Lett. \textbf{85} (2000) 2466.

\bibitem{Schiffer93}
J. P. Schiffer, Phys. Rev. Lett. \textbf{70} (1993)  818.

\bibitem{POSCH}
U. G. Poschinger, et. al., J. Phys. B \textbf{42} (2009) 154013.

\bibitem{NAEGERL98} H. C. N\"agerl, D. Leibfried,
F. Schmidt-Kaler, J. Eschner, R. Blatt, Optics Express \textbf{3}, (1998) 89.

\bibitem{DREWSEN}
M. Drewsen, A. Mortensen, R. Martinussen, P. Staanum,  J. S\"orensen, Phys. Rev. Lett. \textbf{93} (2004) 243201.

\bibitem{POSCH2010} U. G. Poschinger, A. Walther, K. Singer, F. Schmidt-Kaler,
Phys. Rev. Lett. \textbf{105} (2010) 263602.

\bibitem{MYER08}
A. H. Myerson, et. al., Phys. Rev. Lett. \textbf{100} (2008) 200502.

\bibitem{MOEL99}
K. Moelmer, A. Soerensen, Phys. Rev. Lett. \textbf{82} (1999) 1835.





\bibitem{Leib2003} D. Leibfried, et. al., Nature \textbf{422} (2003) 412.

\bibitem{ROOS2008} C. F. Roos, New J. Phys. \textbf{10} (2008) 013002.

\bibitem{KIM2009} K. Kim, et. al., Phys. Rev. Lett. \textbf{103} (2009) 120502.

\bibitem{EDWARDS2010} E. E. Edwards, et. al., Phys. Rev. B \textbf{82} (2010) 060412.

\bibitem{ospelkaus11} C. Ospelkaus, et al.,  arXiv:1104.3573.


\bibitem{Retzker08} A. Retzker, R. C. Thompson, D. M. Segal,  M. B. Plenio,
Phys. Rev. Lett. \textbf{101}(2008) 260504.

\bibitem{Shimshoni11}
E. Shimshoni, G. Morigi, S. Fishman, Phys. Rev. Lett. \textbf{106} (2011) 010401.

\bibitem{ivanov}
P. A. Ivanov, F. Schmidt-Kaler, arXiv:1105.0598.
\bibitem{Schmied09}
R. Schmied, J. H. Wesenberg, D. Leibfried, Phys. Rev. Lett. \textbf{102} (2009) 233002.

\bibitem{kumph}
M. Kumph, M. Brownnutt, R. Blatt, arXiv:1103.5428.

\bibitem{BROWN2011} K. R. Brown, et. al., arXiv:1011.0473.

\bibitem{Meyer} J. S. Meyer, K. A. Matveev, and A. I. Larkin, Phys. Rev. Lett.\textbf{98} (2007) 126404.







\end{thebibliography}
\end{document}